\documentclass[
 aps,
 showpacs,
 showkeys,
 superscriptaddress,
 reprint,%
twocolumn,
]{revtex4-2}
\usepackage{physics}
\usepackage{mathtools}
\usepackage{amssymb, amsfonts, amsthm}
\usepackage{siunitx}
\usepackage[toc,page, titletoc]{appendix}
\usepackage[caption=false]{subfig}
\usepackage{bm}
\usepackage{soul}
\usepackage[dvipsnames]{xcolor}
\usepackage{tikz}
\usetikzlibrary{quantikz}

\usepackage{hyperref}
\hypersetup{colorlinks=true, linkcolor=blue, citecolor=blue, urlcolor=blue, unicode=true}
\usepackage{cleveref}

\newcounter{circuit}[section]

\crefname{circuit}{circuit}{circuits}
\Crefname{circuit}{Circuit}{Circuits}

\newcommand{\Utheta}[1]{U(\theta_{#1})}
\newcommand{\Udagtheta}[1]{U^\dag(\theta_{#1})}
\newcommand{\dtheta}[1]{\partial_{\theta_{#1}}}

\begin{document}

\title{Measurement-induced entanglement phase transitions in variational quantum circuits\\
}

\author{Roeland Wiersema}
\affiliation{Vector Institute, MaRS  Centre,  Toronto,  Ontario,  M5G  1M1,  Canada}
\affiliation{Department of Physics and Astronomy, University of Waterloo, Ontario, N2L 3G1, Canada}

\author{Cunlu Zhou}
\affiliation{Center for Quantum Information and Control, University of New Mexico, Albuquerque, NM, 87131, USA}
\affiliation{Department of Computer Science, University of Toronto, Ontario, M5T 3A1, Canada}
\affiliation{Vector Institute, MaRS  Centre,  Toronto,  Ontario,  M5G  1M1,  Canada}

\author{Juan Felipe Carrasquilla}
\affiliation{Vector Institute, MaRS  Centre,  Toronto,  Ontario,  M5G  1M1,  Canada}
\affiliation{Department of Physics and Astronomy, University of Waterloo, Ontario, N2L 3G1, Canada}
\affiliation{Department of Physics, University of Toronto, Ontario M5S 1A7, Canada}

\author{Yong Baek Kim}
\affiliation{Department of Physics, University of Toronto, Ontario M5S 1A7, Canada}

\date{\today}

\begin{abstract}

Variational quantum algorithms (VQAs), which classically optimize a parametrized quantum circuit to solve a computational task, promise to advance our understanding of quantum many-body systems and improve machine learning algorithms using near-term quantum computers. Prominent challenges associated with this family of quantum-classical hybrid algorithms are the control of quantum entanglement and quantum gradients linked to their classical optimization. Known as the barren plateau phenomenon, these quantum gradients may rapidly vanish in the presence of volume-law entanglement growth, which poses a serious obstacle to the practical utility of VQAs. Inspired by recent studies of measurement-induced entanglement transition in random circuits, we investigate the entanglement transition in variational quantum circuits endowed with intermediate projective measurements. Considering the Hamiltonian Variational Ansatz (HVA) for the XXZ model and the Hardware Efficient Ansatz (HEA), we observe a measurement-induced entanglement transition from volume-law to area-law      with increasing measurement rate. Moreover, we provide evidence that the transition belongs to the same universality class of random unitary circuits. Importantly, the transition coincides with a “landscape transition” from severe to mild/no barren plateaus in the classical optimization. Our work paves an avenue for greatly improving the trainability of quantum circuits by incorporating intermediate measurement protocols in currently available quantum hardware. 

\end{abstract}

\clearpage
\maketitle

Controlling quantum entanglement has been identified as a critical element in the development of quantum computing. A prominent example is variational quantum algorithms, which are designed for quantum simulations and machine learning in near-term quantum computers~\cite{cerezoVariationalQuantumAlgorithms2021}. How to efficiently use quantum entanglement resources impacts the performance of such quantum algorithms. Another important element in a quantum algorithm is the measurement, which can be performed at intermediate steps during a quantum computation and directly affects the overall entanglement structure of the quantum circuit. As there now exists quantum hardware that allows intermediate measurements~\cite{Honeywell,IBM}, it stands to reason that such measurements may offer yet another resource for controlling entanglement and improving VQAs. 

Recently, there has been great progress in understanding the quantum entanglement evolution in random unitary quantum circuits with intermediate projective measurements. In random unitary circuits, the time evolution is governed by unitaries drawn from a random distribution without specifying any Hamiltonian. In these circuits, the nearby two-qubit gates locally entangle qubits, which generally leads to volume-law entanglement growth. When such a system is measured at randomly selected locations throughout the circuit, entanglement is destroyed globally. One might expect that this leads to a simple decrease in the coefficient of the entanglement growth volume law; however, this is not the case. The competition between local entanglement creation and non-local entanglement destruction induces a phase transition in the entanglement growth from a volume to an area law at a critical measurement rate $p_c$~\cite{Chan2019projentdyn, Skinner2019entandynamics, Li2019measdrivqc, Li2018qzeno, cao19entanglementinafermion, Bao2020measurementcrithoneycomb, Czischek2021trappedion, Block2021measind, cao19entanglementinafermion}. Moreover, it appears that this critical behavior is universal, independent of the specific implementation of both the unitary or measurement dynamics. A significant amount of theoretical understanding has been gained about the properties of entanglement phase transitions in random unitary circuits~\cite{Bao2020measurementcrithoneycomb, Jian2020measurementcrithoneycomb} by mapping such systems to well-defined statistical mechanics models. So far, most of the studies on measurement-induced phase transitions are focused on random unitary circuits or circuits with quantum chaotic dynamics. But do these transitions also take place in circuits of practical interest such as the variational quantum circuits used in quantum chemistry, quantum many body simulations, and quantum machine learning?  

In this letter, we show that measurement-induced entanglement phase transitions take place in two prototypical variational quantum circuits, the Hamiltonian variational ansatz (HVA)~\cite{Wecker2015} for the XXZ model and the Hardware efficient ansatz (HEA)~\cite{abhinav17hardware}. 
Notice that the dynamics in the HVA is specified by a Hamiltonian in contrast to random unitaries. The HVA for the XXZ model is of particular interest as it represents the simulation of a non-trivial interacting model of fermions, which is Bethe-ansatz integrable. These quantum circuits are popular examples used within the Variational Quantum Eigensolver (VQE) algorithm~\cite{Peruzzo2014}. This hybrid quantum-classical algorithm is used throughout the literature to approximate quantum many-body ground states~\cite{ho19efficient,Cade2020hvafermi,Wierichs2020avoiding, Wiersema2020exploring, Kattemolle2021kagomehva} or perform quantum chemistry simulations~\cite{abhinav17hardware, Hempel2018qchem, Colless2018qchem, Google2020hartree,OMalley2016qchem}. Our motivations to investigate the measurement-induced entanglement transitions in variational quantum circuits rest on the following two issues. Most of the quantum ground states of interacting many-body systems follow the area-law entanglement (up to a logarithmic correction). However, ballistic growth of entanglement in time evolution implies that circuits used in VQE can rapidly develop much more entanglement than what may be needed to efficiently simulate these ground states of interest~\cite{PhysRevLett.111.127205,PhysRevX.7.031016,Wiersema2020exploring}. 
The second issue is the evaluation of the quantum gradient, which is used to minimize a cost function in VQAs. It is known that quantum circuits that approximate a 2-design have exponentially decaying quantum gradients, localized on so-called barren plateaus, which pose a significant hurdle for variational quantum algorithms~\cite{McClean2018barren, Cerezo2021costfunctiondep, wang2021noiseinduced}. It has also been shown that there is a close relation between entanglement scaling and barren plateaus, hence it is natural to consider constraining the amount of entanglement during parts of the variational optimization as a useful strategy for increasing the trainability of variational circuits~\cite{Marrero2020entanglementbarren,Taylor2020entbarmit,Wiersema2020exploring}. We anticipate that the inclusion of interspersed measurements in the variational quantum circuits may offer an alternative way to control their quantum entanglement, which can be used for more efficient quantum simulations, as well as to overcome the issue of barren plateau in the evaluation of quantum gradients. 

Below we numerically show that the measurement-induced entanglement phase transition in the variational quantum circuits coincides with a “landscape transition”, a change from a landscape with severe barren plateaus to a landscape with mild or no barren plateaus. This suggests that VQE with intermediate projective measurements can potentially be used to avoid barren plateaus and improve current optimization strategies. In deriving our results, we also provide a modified parameter shift rule for calculating the quantum gradients with intermediate projective measurement. 

\begin{figure*}
    \centering
    \includegraphics[width=\textwidth]{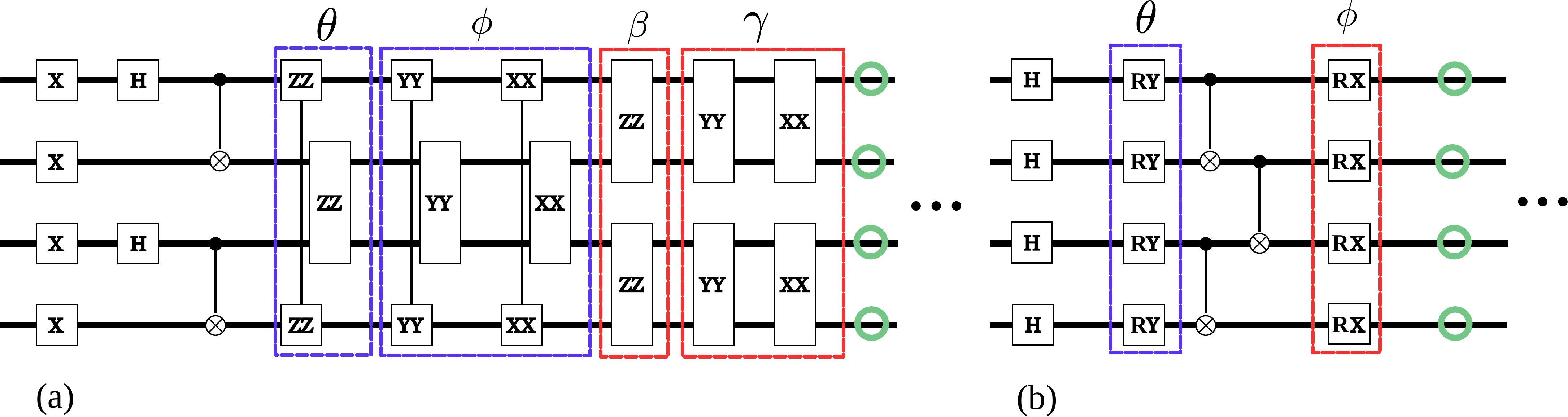}
    \caption{Schematic depiction of the circuits studied in this letter. (a) For the XXZ-HVA, we prepare a Bell state on the even sites and alternatingly apply ZZ, YY and XX two-qubit rotations on odd and even bonds in the chain. For the odd (even) bonds, the ZZ rotations are parameterized by $\theta$ ($\beta$) whereas the YY and XX rotations are parameterized by $\phi$ ($\gamma$). (b)  The initial state in the HEA consists of the equal superposition followed by $L$ layers of low-depth entangling unitaries. These unitaries consists of $N$ Pauli-Y rotations on each qubit, a chain of nearest neighbor CNOTs and $N$ Pauli-X rotations on each qubit. All $2N$ rotations are controlled by individual parameters $\theta_{i,l},\phi_{i,l}$, where $i=1,\ldots,N$ and $l=1,\ldots, L$. After each layer, we perform a projective measurement according to~\cref{eq:single_meas} with probability $p$ on each qubit (indicated by the green circles here), bringing the average number of measurements in the circuits to $NLp$. } 
    \label{fig:circuits}
\end{figure*}

\textit{Measurement--induced entanglement phase transitions}.---We consider a chain of $N$ qubits with spatial periodic boundary conditions.  To quantify the entanglement we consider the bipartite von Neumann entanglement entropy $S(N,p)$ between two halves of an $N$-qubit circuit of depth $L$. At each discrete time step $1\leq d\leq L$, we apply a layer of local unitaries consisting of a combination of single qubit and two-local quantum gates between nearest neighbor qubits, which rapidly increase the entanglement in the chain.  After each layer, we apply projective measurements onto the computational basis $\{\ket{0},\ket{1}\}$ on each qubit with probability $p$. A single measurement on a state $\ket{\psi}$ then results in a state
\begin{align}
    \ket{\psi'} = \frac{\Pi_i \ket{\psi}}{\sqrt{\bra{\psi}\Pi_i \ket{\psi}}} \label{eq:single_meas},
\end{align}
where $\Pi_i = \ketbra{i}$ are the projectors onto $\sigma^z$ basis. These measurements destroy entanglement at any length scale, since the state is locally projected onto a single state~\cite{Li2019measdrivqc}. As a result, the unitary dynamics locally entangles nearest neighbor qubits, whereas measurements globally destroy entanglement between different subsystems. This competition induces a dynamical phase transition between a volume and area law regime of entanglement scaling at a critical measurement rate $p_c$.

Although the critical point $p_c$ can vary between different types of random unitary dynamics and measurement schemes, the critical exponent characterizing the correlation length scale divergence $\xi \propto (p - p_c)^{-\nu}$ appears to be the same for different models at $\nu\approx 4/3$. This critical exponent can be derived by considering toy models and mapping the projective dynamics to a two-dimensional percolation model, which is exactly solvable~\cite{Li2019measdrivqc, Czischek2021trappedion, Jian2020measurementcrithoneycomb, Bao2020measurementcrithoneycomb}. 

Central to the investigations on phase transitions induced by measurements is the concept of steady state entanglement dynamics~\cite{Li2019measdrivqc, Li2018qzeno}. Given a circuit with a number of qubits $N$, we are primarily interested in the late time behavior when $d\to\infty$. In this infinite depth (long time) limit we expect the system to evolve into a steady state, characterized by a typical value  of entanglement entropy that depends on the measurement rate $p$, but not the dynamics at finite times. In order to characterize this regime, we can investigate the average entanglement entropy as a function of depth for different values of $p$. For the moderate system sizes considered in this work, we observe steady state entanglement dynamics at $d=16$. 

\textit{The Variational Quantum Eigensolver}.---Instead of considering Haar random circuits, Floquet dynamics or Random Clifford circuits, we turn to quantum circuits widely used in VQAs, e.g., in the Variational Quantum Eigensolver algorithms~\cite{Peruzzo2014}. This hybrid quantum-classical algorithm takes a quantum circuit $U(\bm{\theta})$ parameterized by a set of parameters $\bm{\theta}$. By invoking the variational principle $E_{\text{ground}} \leq E(\bm{\theta})$, one can use a classical optimization routine to minimize the energy of a Hamiltonian $H$ with respect to the parameterized wave function $\ket{\psi(\bm{\theta})} = U(\bm{\theta})\ket{0}$ and approximate the ground state. As with other variational methods, the choice of ansatz $U(\bm{\theta})$ is crucial since the ground states must be reachable from the initial state by application of this unitary. There exists a variety of proposals, including the so-called Hamiltonian Variational Ansatz (HVA)~\cite{Wecker2015, ho19efficient,Cade2020hvafermi, Wiersema2020exploring, Wierichs2020avoiding, Kattemolle2021kagomehva} and the Hardware efficient Ansatz (HAE)~\cite{abhinav17hardware,Grimsley2019adaptvqe}. The former exploits the structure of the Hamiltonian as the design principle, whereas the latter aims to provide a hardware-friendly parameterization with enough degrees of freedom to capture a variety of states. 

For our numerical study, we investigate the projective dynamics of the XXZ-chain HVA~\cite{ho19efficient,Wierichs2020avoiding,Wiersema2020exploring} and the HEA~\cite{Wecker2015}, whose circuits are depicted in~\cref{fig:circuits}. The former is of particular interest, since the XXZ Hamiltonian describes a system of interacting fermions on a chain, which can be diagonalized via the Bethe ansatz and it is still an open question if Bethe-ansatz integrable models also show measurement-induced entanglement phase transitions~\cite{Bao2020measurementcrithoneycomb}. 

For each circuit, we measure the bipartite entanglement entropy of the output state of the circuit. Note that we are not creating a mixed state by mixing the individual output states into a single density matrix. Instead, we measure the entanglement entropy of a pure state. The average $S(p,N)$ is obtained by averaging over $3 \times 10^3$ circuit realizations with all circuit parameters sampled uniformly in $(0,2\pi)$ and measurements sampled uniformly with probability $p$. Due to the difficulty in simulating large systems, we restrict ourselves to $N = 6,8, \ldots, 18$.

\begin{figure}[htb!]
    \centering
    \subfloat[\label{fig:collapse_hva} XXZ-HVA]{
    \includegraphics[width=0.95\columnwidth]{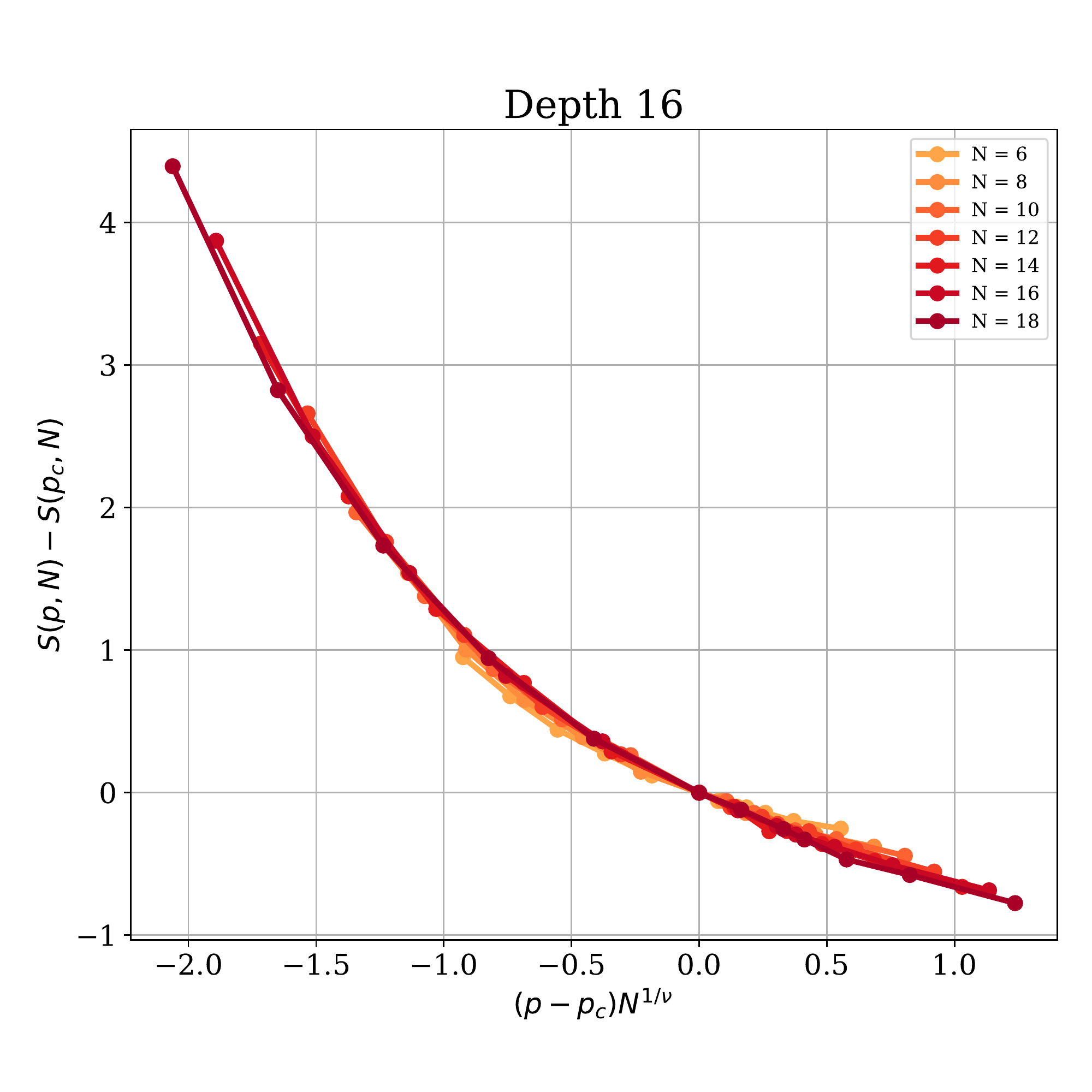}
    }
    
    \subfloat[\label{fig:collapse_haa} HAE]{
\includegraphics[width=0.95\columnwidth]{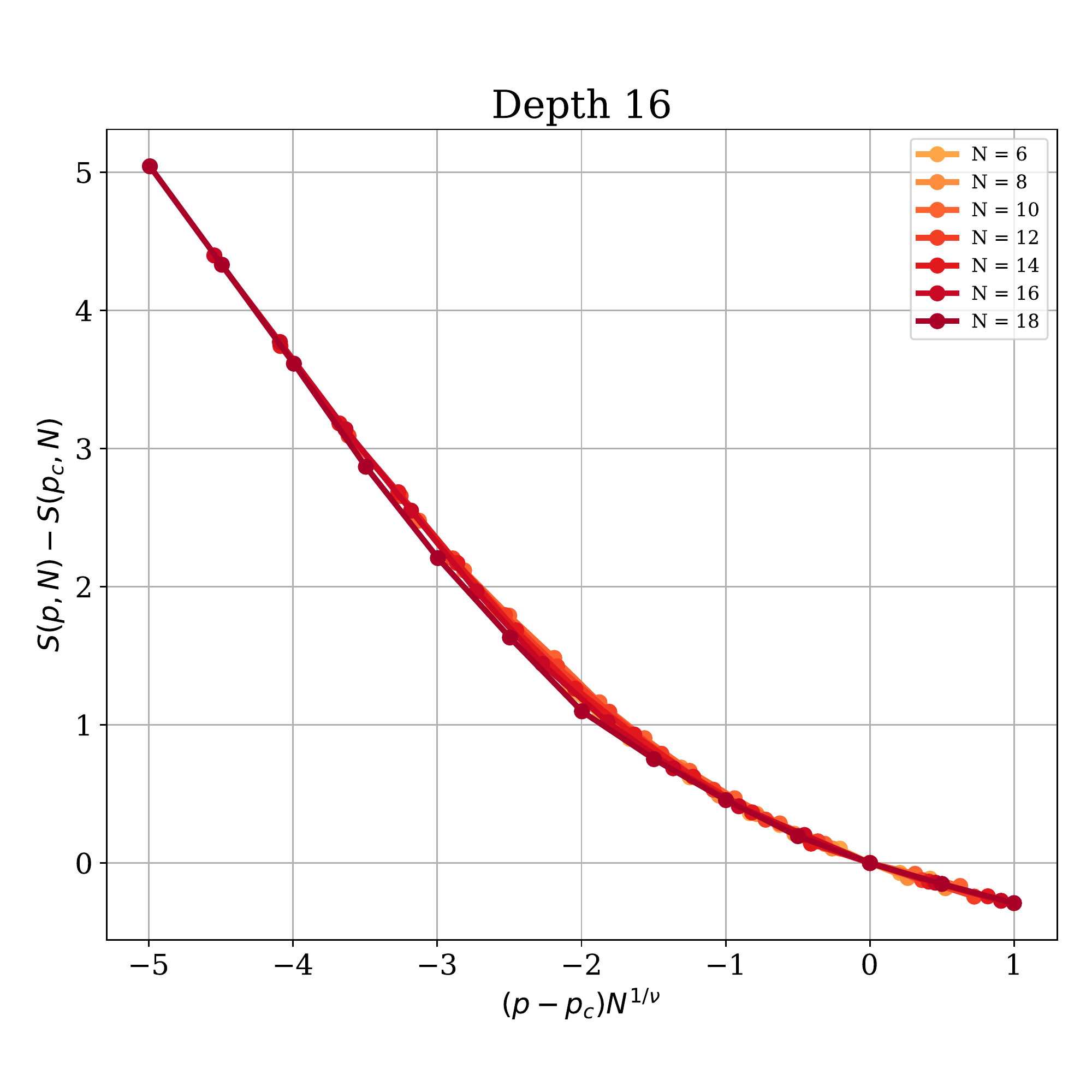}
    }
    \caption{Data collapse of the average entanglement entropies. (a) For the XXZ-HVA, we find $p_c = 0.25$ and $\nu  \approx1.22 \pm 0.24$. (b) For the HEA, we find $p_c\approx 0.5$ and $\nu \approx 1.26 \pm 0.23$. The error bars are calculated as the difference between the critical exponent in the thermodynamic extrapolation and the finite-size data collapse. The entropies are averaged over $3\times 10^3$ circuit realizations.}
    \label{fig:data_collapse}
\end{figure}

\textit{Finite size scaling analysis}.---Since phase transitions only occur in the thermodynamic limit $N\to\infty$, we have to take care of the finite-size effects in analyzing our numerical data. To account for finite-size effects, we fit the scaling form~\cite{Li2019measdrivqc}
\begin{align}
    S(N, p, \nu) - S(N,p_c, \nu)= f(N^{1/ \nu}(p-p_c))
\end{align}
to get a data collapse of the individual circuits of size $N$. To determine $p_c$ and $\nu$, we minimize a Chi-squared statistic between the scaling form above and the data, and use a statistical bootstrap to verify the integrity of the fit. The resulting data collapse can be found in~\cref{fig:data_collapse}. 
To extrapolate the critical exponent to the thermodynamic limit, we do a linear fit of $\nu$ as a function of $1/N'$ where $N_{\max}/2\leq N' \leq N_{\max}$ is the largest value of $N$ in the data set. The intercept then gives us the value of $\nu$ for $N'\to\infty$~\cite{Skinner2019entandynamics}. The details of our statistical estimation procedure are outlined in the supplementary material~\ref{app:finite}.

In addition to the finite scaling analysis, we can investigate the quantum mutual information,
\begin{align}
    I(A,B) = S_A(N,p) + S_B(N,p) - S_{A\cap B}(N,p),
\end{align}
between qubits $A$ and $B$ separated by a distance $r$, which we expect to peak at a critical point due to subsystem correlations becoming non-negligible. From these data, we find similar critical measurement rates $p_c\approx0.25$ and $p_c\approx 0.5$ for the XXZ-HVA and HEA, respectively. In supplementary materials \ref{app:mut_info} we give further details on this procedure.

\textit{Projective gradients and barren plateaus}.---For variational quantum algorithms, recall that one prepares a parameterized quantum circuit (ansatz) on a  quantum computer and feeds the output state to a classical optimizer which updates the parameters by minimizing the energy. Gradient descent is the most widely used classical optimization method for such general non-convex optimization problems. The optimization proceeds by following the opposite direction of the gradient. It has been shown both numerically and analytically that exponentially decaying gradients, localized on so-called barren plateaus, pose a significant hurdle for variational quantum algorithms~\cite{McClean2018barren, Cerezo2021costfunctiondep, wang2021noiseinduced}. As a result, a variety of recent works are aimed at finding ways to circumvent these regions where optimization is hard~\cite{Taylor2020avoidbarren, Volkoff2021avoidbarren, Grant2019initialization, zhou2020qaoa, Cerezo2021costfunctiondep, skolik2020layerwise, pesah2020absence}. Here, we investigate the barren plateaus problems under the influence of projective measurements, more specifically the variance of the gradients in the XXZ-HVA and the HEA with intermediate projective measurements. 

Although, there exists a variety of methods for calculating gradients in quantum circuits~\cite{Mitarai2018, Schuld2019,Wierichs2021gradest, Mari2021gradest, Izmaylov2021gradest,Kyriienko2021gradest}, none of these works consider quantum gradients through a circuit undergoing projective measurements. In~\cite{Koczor2019qngnonuni} the quantum natural gradient~\cite{Stokes2020quantumnatural} is extended to quantum channels. Additionally, in~\cite{Ferguson2020measbasedvqe} measurement-based VQE is investigated, but only in the context of the work by Briegel~\cite{Briegel2009measbased} where an entangled state is prepared and measurement is directly part of the algorithm.

In supplementary materials \ref{app:proj_grads}, we provide a detailed analytical derivation of a quantum gradient in a circuit undergoing $M$ measurements. The result is an estimator for calculating projective gradients 
of the form,
\begin{align}
    \Tr{(\partial_{\theta_l}\rho )O}& = \sum_i^{2^M} \frac{1}{2} \left(\expval{O}^{(i),+} \frac{p^{(i),+}_M}{p^{(i)}_M} - \expval{O}^{(i),-} \frac{p^{(i),-}_M}{p^{(i)}_M}  \right)
    \label{eq:proj_grad}.
\end{align}
Here, the expectation values $\expval{O}^{(i),\pm}$ correspond to the expectation value of $O$ if the circuit has the $\theta_l$ parameter shifted by $\pm \pi/2$ , and the set of outcomes $i=i_1,\ldots i_{M}$ has been observed. The probabilities $p_M^{(i)}$ and $p_M^{(i), \pm}$ are the probabilities of observing outcomes $(i)$. In this work, we have computed those weights exactly and we will explore the scalability of their estimation in future work. 

\begin{figure}[htb!]
    \centering
    \subfloat[\label{fig:grad_var_hva} XXZ-HVA]{
    \includegraphics[width=0.95\columnwidth]{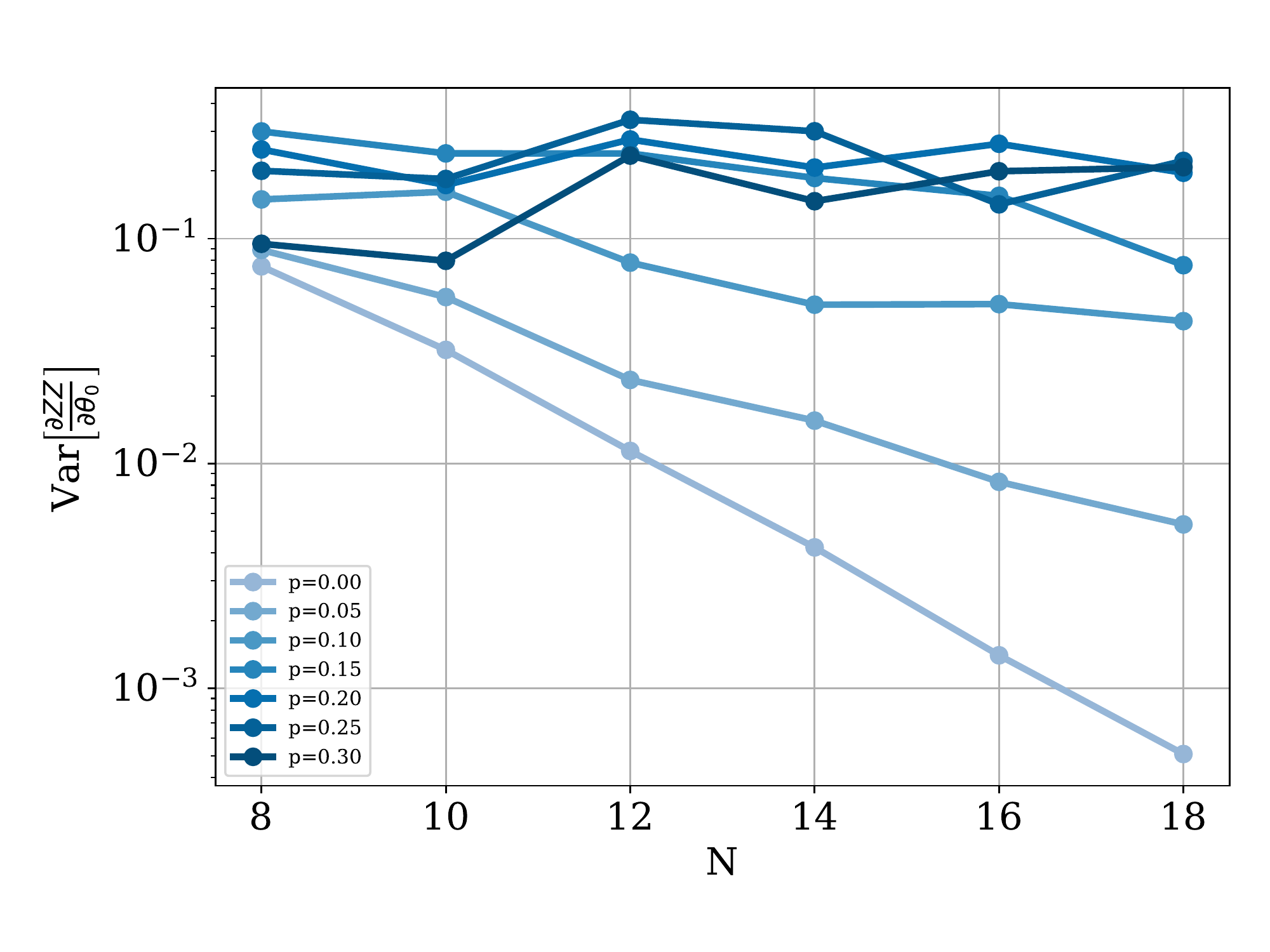}
    }
    
    \subfloat[\label{fig:grad_var_haa}HAE]{
    \includegraphics[width=0.95\columnwidth]{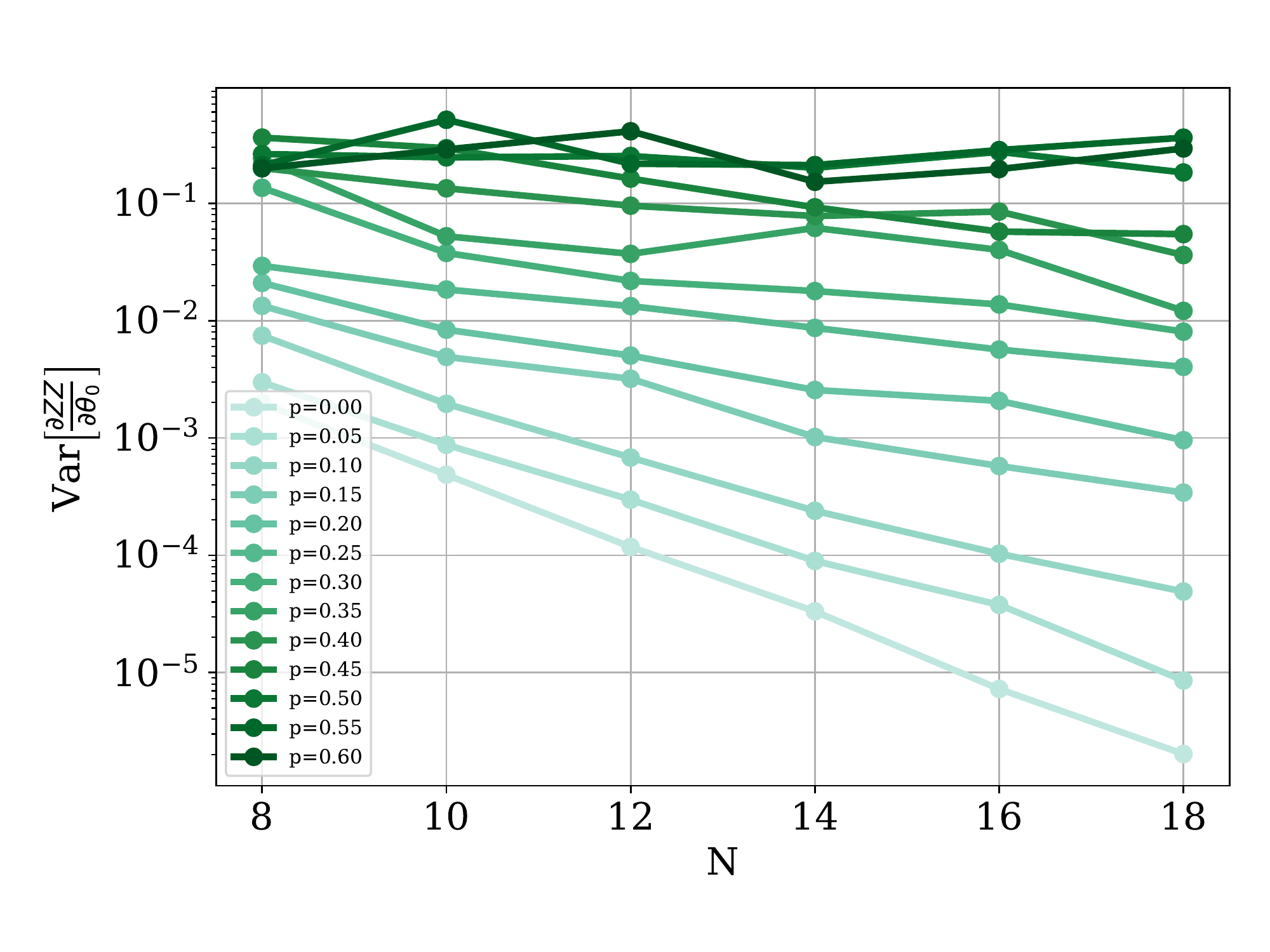}
    }
    \caption{Variance of the projective gradients taken with respect to the first parameter in the circuit ($\theta$ in the first parameterized layer in both the HVA and HEA, see \cref{fig:circuits}). The variances are estimated over $10^3$ samples where for each data point, we randomly choose measurements with probability $p$ and uniformly sample the gate parameters. The gradient is then calculated exactly from \cref{eq:proj_grad}. For the 1D HVA-XXZ circuits with depth $d=16$ (a) and the 1D HAE circuit with depth $d=16$ (b) the gradient variance becomes constant as the measurement rate $p$ increases. The scaling collapse of \cref{eq:grad_var_collapse} gives $\nu\approx 1.31$ and $\nu\approx 1.49$, respectively.}
    \label{fig:grad_var}
\end{figure}

The barren plateau effect takes place in states that have a sufficient amount of randomness \cite{McClean2018barren}. If the states are sampled from the Haar distribution, then we know that the average entanglement entropy is given by the Page entropy \cite{Page1993}, which follows a volume law entanglement scaling. With intermediate projective measurements, volume-law states can be broken down into states with less entanglement, i.e., area-law/logarithmic-law states, and we expect 
that these projective gradients will not suffer from barren plateaus. To investigate this effect, we consider the same circuit settings discussed in \cref{fig:data_collapse} and examine the projective gradients with respect to the expectation value of $H = Z_0 Z_1$. We calculate the projective gradients for the first circuit parameter ($\theta$ in the first parameterized layer in both the HVA and HEA, see \cref{fig:circuits}) using~\cref{eq:proj_grad}. We consider a depth $d=16$ circuit for system sizes $N=8,\ldots 18$. In~\cref{fig:grad_var_hva} and~\cref{fig:grad_var_haa}, we observe that the gradient variances in both the XXZ-HVA and HEA transition from exponentially decaying to a constant as the measurement rate increases. This transition, coincides with the critical measurement rate for the volume-area law transition. Therefore, we see that the measurement-induced entanglement phase transitions induces a landscape transition in the circuit from mild/severe barren plateaus to no barren plateaus. 

Motivated by the observation that the gradient variance decays exponentially for $p<p_c$ and is constant for $p>p_c$, we adopt the following scaling ansatz for the gradient variance:
\begin{align}
    h_\text{var}(N,p, \nu) &= C(N,p_c) + \exp\{-\abs{p-p_c}N^{1/\nu}\}\label{eq:grad_var_collapse}. 
\end{align}
Fitting this function for the previously found critical points gives critical exponents of the same order as found previously, although due to the noise in the data the results are not as reliable as the data collapse of~\cref{fig:data_collapse}. 

\textit{Outlook}.---In this work, we demonstrated the existence of a measurement-induced entanglement transition in variational quantum circuits which coincides with a “landscape transition” in the behavior of quantum gradients. As mentioned earlier, the exponentially-vanishing quantum gradients in presence of volume-law entanglement growth, the so-called barren plateau, is a serious obstacle in the applications of variational quantum circuits. Our work suggests that the intermediate projective measurements may provide a useful knob to control the barren plateau issue. Inclusion of the measurement protocol in the quantum-classical hybrid algorithm would be a timely development given that quantum computing hardware companies like IBM and Honeywell now allow their users to perform mid-circuit measurements, enabling the real-time logic required for performing these algorithms in an experimental setting~\cite{Honeywell, IBM}. In particular, the Hamiltonian variational quantum circuits considered in this work could be easily implemented in the quantum hardware. A detailed analysis of when and how a projective circuit optimization can be “advantageous” would be an excellent topic of future study.

For practical implementation of the measurement protocol, note that the scheme we provided here is quite general and many extensions and modifications are possible. For instance, the projective measurements used in this work can be replaced by general Positive Operator Value Measures (POVM) or parameterized measurements. Additionally, we have focused on one-dimensional quantum circuits where the measurement-induced entanglement transition belongs to the same universality class as in the random unitary circuits. It would be interesting to consider moderately sized quantum circuits with a two-dimensional topology, and see if a similar phase transition appears there and investigate the universality class. 

\textit{Acknowledgements}.---We would like to thank Henry Yuen for the discussions during the course of this project. Y.B.K. is supported by the NSERC of Canada and the Center for Quantum Materials at the University of Toronto. J.C. acknowledges support from NSERC, the Shared Hierarchical Academic Research Computing Network (SHARCNET), Compute Canada, Google Quantum Research Award, and the CIFAR AI chair program. Resources used in preparing this research were provided, in part, by the Province of Ontario, the Government of Canada through CIFAR, and companies sponsoring the Vector Institute \url{www.vectorinstitute.ai/#partners}. C.Z. acknowledges support from the U.S. National Science Foundation under Grant No. 2116246 and the U.S. Department of Energy, Office of Science, National Quantum Information Science Research Centers, Quantum Systems Accelerator, and the Postgraduate Affiliate Award from the Vector Institute. 
\bibliographystyle{unsrt}
\bibliography{library.bib}

\begin{thebibliography}{10}

\bibitem{cerezoVariationalQuantumAlgorithms2021}
M.~Cerezo, Andrew Arrasmith, Ryan Babbush, Simon~C. Benjamin, Suguru Endo,
  Keisuke Fujii, Jarrod~R. McClean, Kosuke Mitarai, Xiao Yuan, Lukasz Cincio,
  and Patrick~J. Coles.
\newblock Variational quantum algorithms.
\newblock {\em Nature Reviews Physics}, 3(9):625--644, September 2021.

\bibitem{Honeywell}
Honeywell Quantum.
\newblock {Mid-circuit measurements on the System Model H1}.
\newblock
  \url{https://www.honeywell.com/us/en/company/quantum/quantum-computer/},
  2021.
\newblock [Online; accessed 07/18/2021].

\bibitem{IBM}
IBM Quantum.
\newblock {Mid-circuit Measurements Tutorial}.
\newblock
  \url{https://quantum-computing.ibm.com/lab/docs/iql/manage/systems/midcircuit-measurement/},
  2021.
\newblock [Online; accessed 07/18/2021].

\bibitem{Chan2019projentdyn}
Amos Chan, Rahul~M. Nandkishore, Michael Pretko, and Graeme Smith.
\newblock Unitary-projective entanglement dynamics.
\newblock {\em Phys. Rev. B}, 99:224307, 6 2019.

\bibitem{Skinner2019entandynamics}
Brian Skinner, Jonathan Ruhman, and Adam Nahum.
\newblock Measurement-induced phase transitions in the dynamics of
  entanglement.
\newblock {\em Phys. Rev. X}, 9:031009, 07 2019.

\bibitem{Li2019measdrivqc}
Yaodong Li, Xiao Chen, and Matthew P.~A. Fisher.
\newblock Measurement-driven entanglement transition in hybrid quantum
  circuits.
\newblock {\em Phys. Rev. B}, 100:134306, 10 2019.

\bibitem{Li2018qzeno}
Yaodong Li, Xiao Chen, and Matthew P.~A. Fisher.
\newblock Quantum zeno effect and the many-body entanglement transition.
\newblock {\em Phys. Rev. B}, 98:205136, 11 2018.

\bibitem{cao19entanglementinafermion}
Xiangyu Cao, Antoine Tilloy, and Andrea~De Luca.
\newblock {Entanglement in a fermion chain under continuous monitoring}.
\newblock {\em SciPost Phys.}, 7:24, 2019.

\bibitem{Bao2020measurementcrithoneycomb}
Yimu Bao, Soonwon Choi, and Ehud Altman.
\newblock Theory of the phase transition in random unitary circuits with
  measurements.
\newblock {\em Phys. Rev. B}, 101:104301, 3 2020.

\bibitem{Czischek2021trappedion}
Stefanie Czischek, Giacomo Torlai, Sayonee Ray, Rajibul Islam, and Roger~G.
  Melko.
\newblock Simulating a measurement-induced phase transition for trapped ion
  circuits, 2021.

\bibitem{Block2021measind}
Maxwell Block, Yimu Bao, Soonwon Choi, Ehud Altman, and Norman Yao.
\newblock The measurement-induced transition in long-range interacting quantum
  circuits, 2021.

\bibitem{Jian2020measurementcrithoneycomb}
Chao-Ming Jian, Yi-Zhuang You, Romain Vasseur, and Andreas W.~W. Ludwig.
\newblock Measurement-induced criticality in random quantum circuits.
\newblock {\em Phys. Rev. B}, 101:104302, 3 2020.

\bibitem{Wecker2015}
Dave Wecker, Matthew~B. Hastings, and Matthias Troyer.
\newblock Progress towards practical quantum variational algorithms.
\newblock {\em Phys. Rev. A}, 92:042303, 2015.

\bibitem{abhinav17hardware}
Abhinav Kandala, Antonio Mezzacapo, Kristan Temme, Maika Takita, Markus Brink,
  Jerry~M. Chow, and Jay~M. Gambetta.
\newblock Hardware-efficient variational quantum eigensolver for small
  molecules and quantum magnets.
\newblock {\em Nature}, 549(7671):242--246, 2017.

\bibitem{Peruzzo2014}
Alberto Peruzzo, Jarrod McClean, Peter Shadbolt, Man-Hong Yung, Xiao-Qi Zhou,
  Peter~J. Love, Al{\'a}n Aspuru-Guzik, and Jeremy~L. O'Brien.
\newblock A variational eigenvalue solver on a photonic quantum processor.
\newblock {\em Nature Communications}, 5(1):4213, 2014.

\bibitem{ho19efficient}
Wen~Wei Ho and Timothy~H. Hsieh.
\newblock {Efficient variational simulation of non-trivial quantum states}.
\newblock {\em SciPost Phys.}, 6:29, 2019.

\bibitem{Cade2020hvafermi}
Chris Cade, Lana Mineh, Ashley Montanaro, and Stasja Stanisic.
\newblock Strategies for solving the fermi-hubbard model on near-term quantum
  computers.
\newblock {\em Phys. Rev. B}, 102:235122, Dec 2020.

\bibitem{Wierichs2020avoiding}
David Wierichs, Christian Gogolin, and Michael Kastoryano.
\newblock Avoiding local minima in variational quantum eigensolvers with the
  natural gradient optimizer.
\newblock {\em Phys. Rev. Research}, 2:043246, Nov 2020.

\bibitem{Wiersema2020exploring}
Roeland Wiersema, Cunlu Zhou, Yvette de~Sereville, Juan~Felipe Carrasquilla,
  Yong~Baek Kim, and Henry Yuen.
\newblock Exploring entanglement and optimization within the hamiltonian
  variational ansatz.
\newblock {\em PRX Quantum}, 1:020319, 12 2020.

\bibitem{Kattemolle2021kagomehva}
Joris Kattem{\"o}lle and Jasper van Wezel.
\newblock Variational quantum eigensolver for the heisenberg antiferromagnet on
  the kagome lattice, 2021.

\bibitem{Hempel2018qchem}
Cornelius Hempel, Christine Maier, Jonathan Romero, Jarrod McClean, Thomas
  Monz, Heng Shen, Petar Jurcevic, Ben~P. Lanyon, Peter Love, Ryan Babbush,
  Al\'an Aspuru-Guzik, Rainer Blatt, and Christian~F. Roos.
\newblock Quantum chemistry calculations on a trapped-ion quantum simulator.
\newblock {\em Phys. Rev. X}, 8:031022, Jul 2018.

\bibitem{Colless2018qchem}
J.~I. Colless, V.~V. Ramasesh, D.~Dahlen, M.~S. Blok, M.~E. Kimchi-Schwartz,
  J.~R. McClean, J.~Carter, W.~A. de~Jong, and I.~Siddiqi.
\newblock Computation of molecular spectra on a quantum processor with an
  error-resilient algorithm.
\newblock {\em Phys. Rev. X}, 8:011021, Feb 2018.

\bibitem{Google2020hartree}
null null, Frank Arute, Kunal Arya, Ryan Babbush, Dave Bacon, Joseph~C. Bardin,
  Rami Barends, Sergio Boixo, Michael Broughton, Bob~B. Buckley, David~A.
  Buell, Brian Burkett, Nicholas Bushnell, Yu~Chen, Zijun Chen, Benjamin
  Chiaro, Roberto Collins, William Courtney, Sean Demura, Andrew Dunsworth,
  Edward Farhi, Austin Fowler, Brooks Foxen, Craig Gidney, Marissa Giustina,
  Rob Graff, Steve Habegger, Matthew~P. Harrigan, Alan Ho, Sabrina Hong, Trent
  Huang, William~J. Huggins, Lev Ioffe, Sergei~V. Isakov, Evan Jeffrey, Zhang
  Jiang, Cody Jones, Dvir Kafri, Kostyantyn Kechedzhi, Julian Kelly, Seon Kim,
  Paul~V. Klimov, Alexander Korotkov, Fedor Kostritsa, David Landhuis, Pavel
  Laptev, Mike Lindmark, Erik Lucero, Orion Martin, John~M. Martinis, Jarrod~R.
  McClean, Matt McEwen, Anthony Megrant, Xiao Mi, Masoud Mohseni, Wojciech
  Mruczkiewicz, Josh Mutus, Ofer Naaman, Matthew Neeley, Charles Neill, Hartmut
  Neven, Murphy~Yuezhen Niu, Thomas~E. O’Brien, Eric Ostby, Andre Petukhov,
  Harald Putterman, Chris Quintana, Pedram Roushan, Nicholas~C. Rubin, Daniel
  Sank, Kevin~J. Satzinger, Vadim Smelyanskiy, Doug Strain, Kevin~J. Sung,
  Marco Szalay, Tyler~Y. Takeshita, Amit Vainsencher, Theodore White, Nathan
  Wiebe, Z.~Jamie Yao, Ping Yeh, and Adam Zalcman.
\newblock Hartree-fock on a superconducting qubit quantum computer.
\newblock {\em Science}, 369(6507):1084--1089, 2020.

\bibitem{OMalley2016qchem}
P.~J.~J. O'Malley, R.~Babbush, I.~D. Kivlichan, J.~Romero, J.~R. McClean,
  R.~Barends, J.~Kelly, P.~Roushan, A.~Tranter, N.~Ding, B.~Campbell, Y.~Chen,
  Z.~Chen, B.~Chiaro, A.~Dunsworth, A.~G. Fowler, E.~Jeffrey, E.~Lucero,
  A.~Megrant, J.~Y. Mutus, M.~Neeley, C.~Neill, C.~Quintana, D.~Sank,
  A.~Vainsencher, J.~Wenner, T.~C. White, P.~V. Coveney, P.~J. Love, H.~Neven,
  A.~Aspuru-Guzik, and J.~M. Martinis.
\newblock Scalable quantum simulation of molecular energies.
\newblock {\em Phys. Rev. X}, 6:031007, Jul 2016.

\bibitem{PhysRevLett.111.127205}
Hyungwon Kim and David~A. Huse.
\newblock Ballistic spreading of entanglement in a diffusive nonintegrable
  system.
\newblock {\em Phys. Rev. Lett.}, 111:127205, Sep 2013.

\bibitem{PhysRevX.7.031016}
Adam Nahum, Jonathan Ruhman, Sagar Vijay, and Jeongwan Haah.
\newblock Quantum entanglement growth under random unitary dynamics.
\newblock {\em Phys. Rev. X}, 7:031016, Jul 2017.

\bibitem{McClean2018barren}
Jarrod~R. McClean, Sergio Boixo, Vadim~N. Smelyanskiy, Ryan Babbush, and
  Hartmut Neven.
\newblock Barren plateaus in quantum neural network training landscapes.
\newblock {\em Nature Communications}, 9(1):4812, Nov 2018.

\bibitem{Cerezo2021costfunctiondep}
M.~Cerezo, Akira Sone, Tyler Volkoff, Lukasz Cincio, and Patrick~J. Coles.
\newblock Cost function dependent barren plateaus in shallow parametrized
  quantum circuits.
\newblock {\em Nature Communications}, 12(1):1791, Mar 2021.

\bibitem{wang2021noiseinduced}
Samson Wang, Enrico Fontana, M.~Cerezo, Kunal Sharma, Akira Sone, Lukasz
  Cincio, and Patrick~J. Coles.
\newblock Noise-induced barren plateaus in variational quantum algorithms,
  2021.

\bibitem{Marrero2020entanglementbarren}
Carlos~Ortiz Marrero, Mária Kieferová, and Nathan Wiebe.
\newblock Entanglement induced barren plateaus, 2020.

\bibitem{Taylor2020entbarmit}
Taylor~L. Patti, Khadijeh Najafi, Xun Gao, and Susanne~F. Yelin.
\newblock Entanglement devised barren plateau mitigation, 2020.

\bibitem{Grimsley2019adaptvqe}
Harper~R. Grimsley, Sophia~E. Economou, Edwin Barnes, and Nicholas~J. Mayhall.
\newblock An adaptive variational algorithm for exact molecular simulations on
  a quantum computer.
\newblock {\em Nature Communications}, 10(1):3007, Jul 2019.

\bibitem{Taylor2020avoidbarren}
Taylor~L. Patti, Khadijeh Najafi, Xun Gao, and Susanne~F. Yelin.
\newblock Entanglement devised barren plateau mitigation, 2020.

\bibitem{Volkoff2021avoidbarren}
Tyler Volkoff and Patrick~J Coles.
\newblock Large gradients via correlation in random parameterized quantum
  circuits.
\newblock {\em Quantum Science and Technology}, 6(2):025008, jan 2021.

\bibitem{Grant2019initialization}
Edward Grant, Leonard Wossnig, Mateusz Ostaszewski, and Marcello Benedetti.
\newblock An initialization strategy for addressing barren plateaus in
  parametrized quantum circuits.
\newblock {\em {Quantum}}, 3:214, December 2019.

\bibitem{zhou2020qaoa}
Leo Zhou, Sheng-Tao Wang, Soonwon Choi, Hannes Pichler, and Mikhail~D. Lukin.
\newblock Quantum approximate optimization algorithm: Performance, mechanism,
  and implementation on near-term devices.
\newblock {\em Phys. Rev. X}, 10:021067, Jun 2020.

\bibitem{skolik2020layerwise}
Andrea Skolik, Jarrod~R. McClean, Masoud Mohseni, Patrick van~der Smagt, and
  Martin Leib.
\newblock Layerwise learning for quantum neural networks, 2020.

\bibitem{pesah2020absence}
Arthur Pesah, M.~Cerezo, Samson Wang, Tyler Volkoff, Andrew~T. Sornborger, and
  Patrick~J. Coles.
\newblock Absence of barren plateaus in quantum convolutional neural networks,
  2020.

\bibitem{Mitarai2018}
K.~Mitarai, M.~Negoro, M.~Kitagawa, and K.~Fujii.
\newblock Quantum circuit learning.
\newblock {\em Phys. Rev. A}, 98:032309, 2018.

\bibitem{Schuld2019}
Maria Schuld, Ville Bergholm, Christian Gogolin, Josh Izaac, and Nathan
  Killoran.
\newblock Evaluating analytic gradients on quantum hardware.
\newblock {\em Phys. Rev. A}, 99:032331, 2019.

\bibitem{Wierichs2021gradest}
David Wierichs, Josh Izaac, Cody Wang, and Cedric Yen-Yu Lin.
\newblock General parameter-shift rules for quantum gradients, 2021.

\bibitem{Mari2021gradest}
Andrea Mari, Thomas~R. Bromley, and Nathan Killoran.
\newblock Estimating the gradient and higher-order derivatives on quantum
  hardware.
\newblock {\em Phys. Rev. A}, 103:012405, Jan 2021.

\bibitem{Izmaylov2021gradest}
Artur~F. Izmaylov, Robert~A. Lang, and Tzu-Ching Yen.
\newblock Analytic gradients in variational quantum algorithms: Algebraic
  extensions of the parameter-shift rule to general unitary transformations,
  2021.

\bibitem{Kyriienko2021gradest}
Oleksandr Kyriienko and Vincent~E. Elfving.
\newblock Generalized quantum circuit differentiation rules, 2021.

\bibitem{Koczor2019qngnonuni}
Bálint Koczor and Simon~C. Benjamin.
\newblock Quantum natural gradient generalised to non-unitary circuits, 2019.

\bibitem{Stokes2020quantumnatural}
James Stokes, Josh Izaac, Nathan Killoran, and Giuseppe Carleo.
\newblock {Quantum {N}atural {G}radient}.
\newblock {\em {Quantum}}, 4:269, May 2020.

\bibitem{Ferguson2020measbasedvqe}
Ryan~R. Ferguson, Luca Dellantonio, Karl Jansen, Abdulrahim~Al Balushi,
  Wolfgang Dür, and Christine~A. Muschik.
\newblock A measurement-based variational quantum eigensolver, 2020.

\bibitem{Briegel2009measbased}
H.~J. Briegel, D.~E. Browne, W.~D{\"u}r, R.~Raussendorf, and M.~Van~den Nest.
\newblock Measurement-based quantum computation.
\newblock {\em Nature Physics}, 5(1):19--26, Jan 2009.

\bibitem{Page1993}
Don~N. Page.
\newblock Average entropy of a subsystem.
\newblock {\em Phys. Rev. Lett.}, 71:1291--1294, 1993.

\bibitem{Newman99montecarlofinite}
M.~E.~J. Newman and G.~T. Barkema.
\newblock {\em Monte Carlo methods in statistical physics}.
\newblock Clarendon Press, Oxford, 1999.

\bibitem{Nelder1965gaussian}
J.~A. Nelder and R.~Mead.
\newblock {A Simplex Method for Function Minimization}.
\newblock {\em The Computer Journal}, 7(4):308--313, 01 1965.

\end{thebibliography}

\clearpage
\onecolumngrid
\renewcommand{\appendixtocname}{Supplementary material}

\renewcommand{\thesection}{\Alph{section}}
\setcounter{section}{0}
\pagebreak
\widetext
\begin{center}
\textbf{\large Supplemental Materials}
\end{center}
\setcounter{equation}{0}
\setcounter{figure}{0}
\setcounter{table}{0}
\setcounter{page}{1}
\makeatletter

\section{Finite-scaling analysis and data collapse \label{app:finite}}

The correlation length $\xi$ of a system quantifies the length scale over which parts of a system are correlated. When a system undergoes a phase transition, information has to propagate throughout the entire system to induce a change of the physical properties, and as a result, the correlation length diverges. Phase transitions only occur in the thermodynamic limit, and hence simulations of finite-sized systems will contain artifacts that have to be accounted for in order to capture the correct behavior~\cite{Newman99montecarlofinite}. In particular, for a finite system the correlation length $\xi$ cannot become infinite and is cut off at $L^d$, the maximum volume of a finite $d$-dimensional system. 
To account for this effect, we can perform a finite-scaling analysis.

The entanglement entropy as a function of measurement rate is conjectured to follow a volume law for $p<p_c$, a constant plus logarithmic correction at $p = p_c$ and area law for $p>p_c$~\cite{Skinner2019entandynamics, Li2019measdrivqc, Bao2020measurementcrithoneycomb}. We can therefore construct a scaling form of the entanglement entropy as
\begin{align}
    S(N, p, \nu) =  S(N,p_c, \nu)+ f(N^{1/ \nu}(p-p_c)) \label{eq:scaling_S}
\end{align}
where $S(N, p, \nu)$ denotes the von Neumann entropy at measurement rate $p$ and $f$ is a scaling function. The critical exponent $\nu$ determines the scaling of the entanglement entropy near $p_c$. If this scaling form is correct, we should be able to account for finite-size effects and all the data can be appropriately rescaled to match a single curve representing $f$ with a proper choice of $\nu$. 

To determine the critical exponents, we fit a 5th-degree polynomial $g$ to our data using a Nelder-Mead optimization~\cite{Nelder1965gaussian} and minimize the $\chi^2$-statistic
\begin{align}
    \chi^2 = \sum_i \frac{(S(N_i, p_i, \nu) - \tilde{S}(N_i,p_i,\nu))^2}{\Delta S}.
\end{align}
Here, $\tilde{S}(N_i,p_i,\nu)$ is estimated from the data and $S(N_i, p_i, \nu)$ is the proposed scaling form from \cref{eq:scaling_S}.
$\Delta S$ is the standard deviation of the von Neumann entropies which arises due to the fluctuations induced by the randomized measurements and their outcomes. From the unscaled data, we determine a set of potential critical points $p_c$ and fit the above $\chi^2$-statistic to determine $\nu$. We then report the values of $p_c$ and $\nu$ that provided the best fit.

To verify the stability of the fit, we perform a statistical bootstrapping procedure to estimate the error bars on the fitted critical exponent $\nu$. We take $K_{\text{boot}} = 100$, where each data set consists of $K$ samples obtained by sampling from the entire data set of $3\times 10^3$ data points with replacement. The final obtained error bars on $\nu$ are $\approx 0.01$.

We can extrapolate our result to the thermodynamic limit by fitting the data for $N' = N_{\text{max}} / 2$ to $N' = N_{\text{max}}$ and plotting the resulting values for $\nu$ against $1/N'$~\cite{Skinner2019entandynamics}. By doing a linear fit on the resulting data, we obtain
\begin{align}
    \tilde{\nu}(N') = a \frac{1}{N'} + b
\end{align}
and so the intercept $b$ corresponds to the value of $\nu$ in the thermodynamic limit, since $\lim_{N'\to\infty} 1/N' = 0$. When fitting the data, we weigh the errors by the standard errors obtained in the statistical bootstrap described above.
\clearpage
\section{Mutual information\label{app:mut_info}}
\begin{figure}[htb!]
    \centering
    \subfloat[\label{fig:mut_info_hva} XXZ-HVA]{
    \includegraphics[width=0.5\columnwidth]{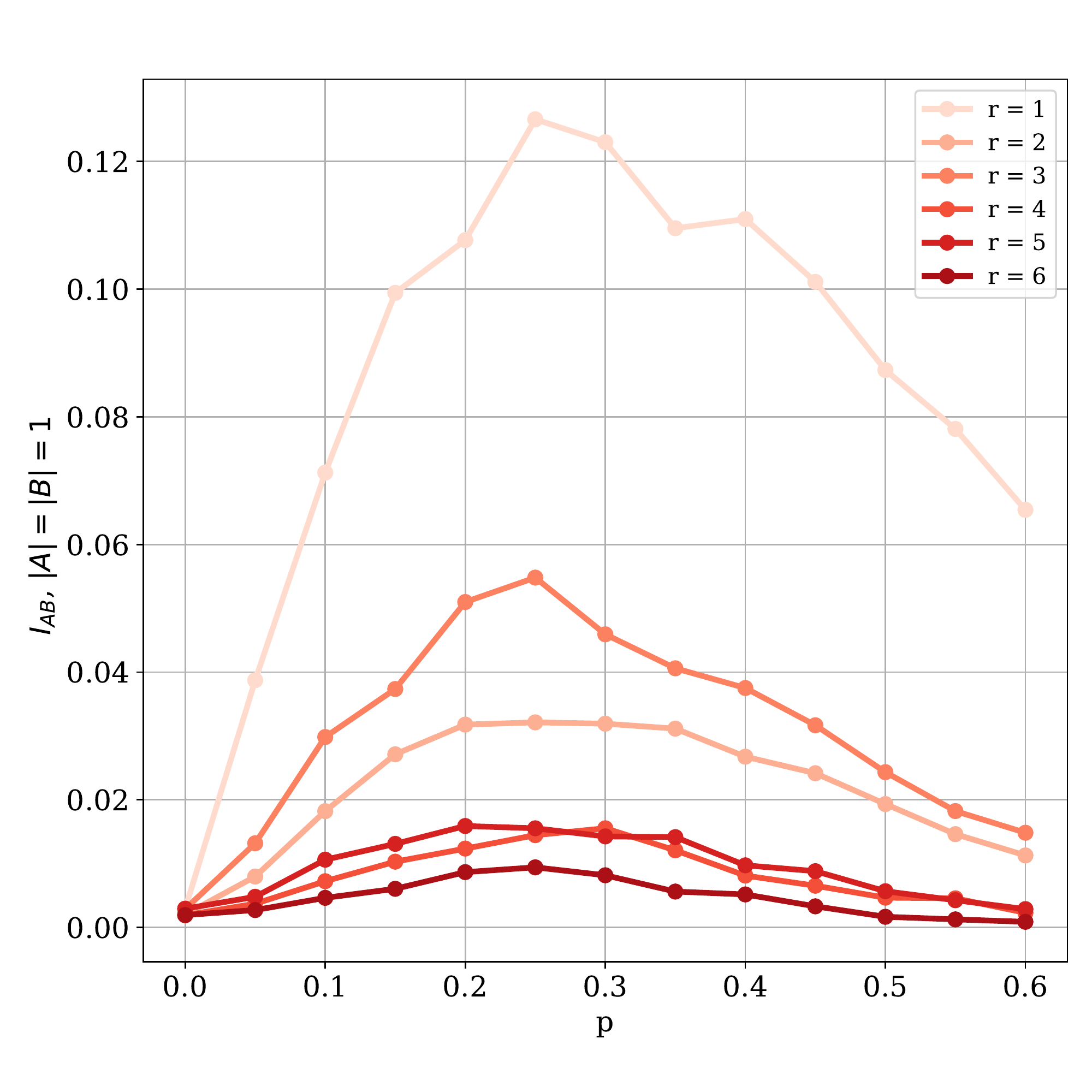}
    }
    \subfloat[\label{fig:mut_info_haa}HAE]{
    \includegraphics[width=0.5\columnwidth]{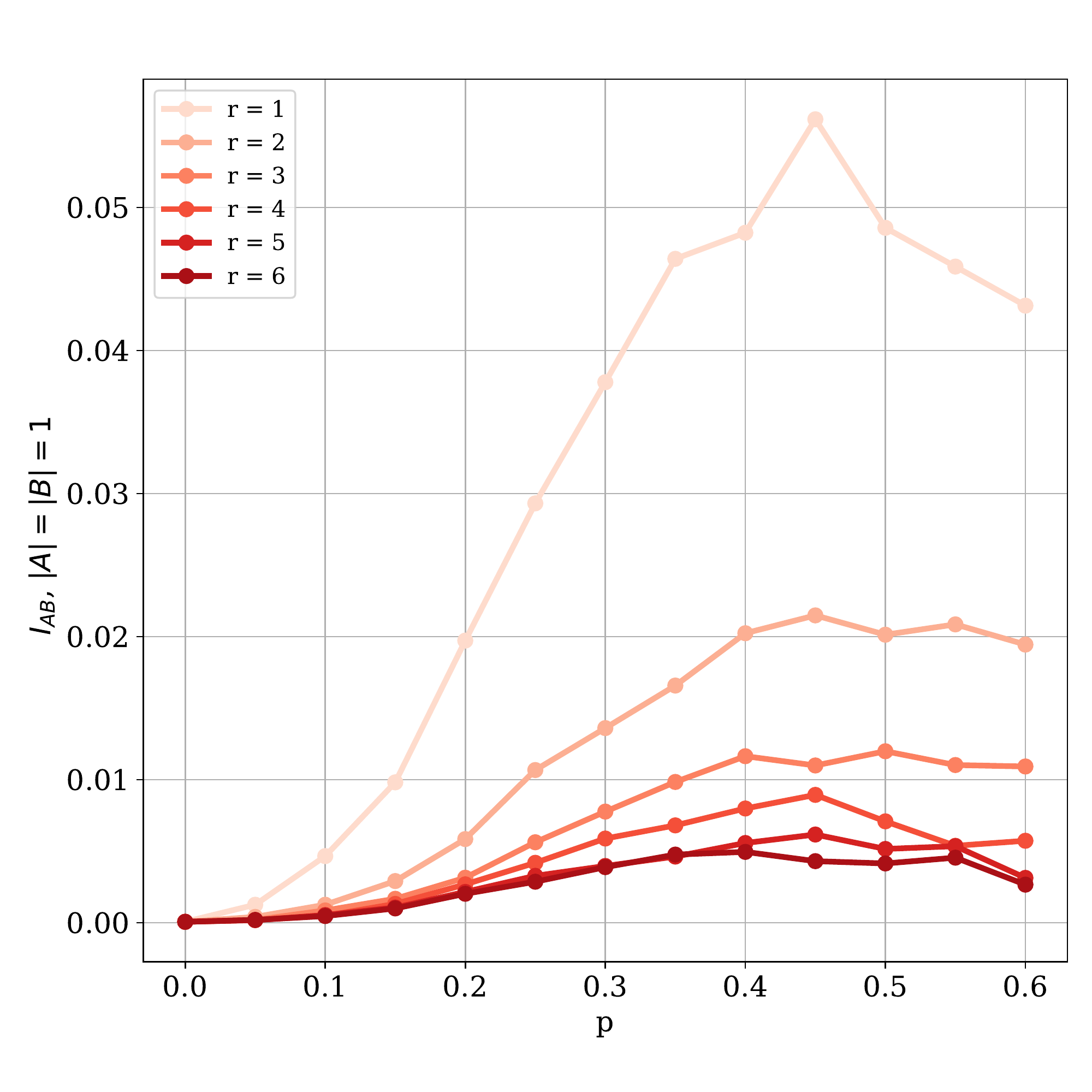}
    }
    \caption{Quantum mutual information between two qubits $A$ and $B$ separated by a distance $r$ on a chain of length $16$. The mutual information is averaged over $3 \times 10^3$ samples, where each sample corresponds to a random circuit realization, as described in the main text.}
    \label{fig:mut_info}
\end{figure}

The quantum mutual information can be used to quantify subsystem correlations, and subsequently detect phase transitions since we expect correlations to divergence at criticality ~\cite{Skinner2019entandynamics, Li2019measdrivqc, cao19entanglementinafermion}. As additional confirmation that the critical values $p_c$ estimated from the prior analysis are correct, we calculate the quantum mutual information as,
\begin{align}
    I(A,B) = S_A(N,p) + S_B(N,p) - S_{A\cap B}(N,p)
\end{align}
Here, we take the same approach as in \cite{Li2019measdrivqc}, and take $A$ and $B$ to be two single qubit subsystems $\abs{A}=1$ and $\abs{B}=1$. We then vary the distance $r$ between qubit $A$ and $B$, to determine the effect of the distance on the subsystem correlations. In \cref{fig:mut_info}, we observe two broad peaks around the previously found values $p_c\approx0.25$ and $p_c\approx0.5$ for the XXZ-HVA and HAA, respectively. 

\clearpage

\section{Projective gradients \label{app:proj_grads}}

Let $\ket{\psi}$ be a quantum state of an $n$-qubit system with corresponding density operator $ \ketbra{\psi}\rho\in L(\mathbb{C}^{2^n})$. A projective measurement can transform the state as
\begin{align}
    \ket{\psi} &\mapsto \frac{\Pi \ket{\psi}}{\bra{\psi}\Pi \ket{\psi}},\\
    \rho &\mapsto \frac{\Pi \rho \Pi}{ \Tr{\Pi \rho}},
\end{align}
where $\Pi$ is a projector onto an eigenbasis of some Hermitian observable $O$ and is therefore Hermitian itself. Since $\Pi$ is a projector it satisfies $\Pi^2 = \Pi$. The normalization constant $\Tr{\Pi\rho}$ gives the overlap of the state $\rho$ with the basis onto which $\Pi$ projects the state. We can write a parameterized state as 
\begin{align}
    \ket{\psi(\theta_{1:k})} = \overleftarrow{\prod_{k=1}} U(\theta_k)\ket{\psi_0},
\end{align}
where we use the shorthand notation $\theta_{1:k} \equiv (\theta_1,\ldots,\theta_k)$ and $\overleftarrow{\prod}$ indicates that the product is ordered from right to left.

We construct the gradient update rule inductively. First, we calculate the resulting state for 1, 2 and $M$ measurements applied to the unitary circuit. Consider an initial state $\rho_0 = \ketbra{\psi_0}$, to which we apply a unitary matrix $\Utheta{1}$ followed by a projective measurement $\Pi_1$,
\begin{align}
    \rho_1(\theta_1) &= \frac{\Pi_1 \Utheta{1}\rho_0 \Udagtheta{1} \Pi_1}{\Tr{\Pi_1 \Utheta{1}\rho_0 \Udagtheta{1}\Pi_1}} \\
    &= \frac{\Pi_1 \Utheta{1}\rho_0 \Udagtheta{1} \Pi_1}{p_1(\theta_1)} .
\end{align}
Next, we add an additional unitary and measurement,
\begin{align}
    \rho_2(\theta_1, \theta_2) & = \frac{\Pi_2 \Utheta{2} \rho_1(\theta_1) \Udagtheta{2} \Pi_2}{\Tr{\Pi_2 \Utheta{2} \rho_1(\theta_1) \Udagtheta{2} \Pi_2}} \\
    & = \frac{\Pi_2 \Utheta{2} \Pi_1 \Utheta{1}\rho_0 \Udagtheta{1} \Pi_1 \Udagtheta{2} \Pi_2}{\Tr{\Pi_2 \Utheta{2} \Pi_1 \Utheta{1}\rho_0 \Udagtheta{1} \Pi_1 \Udagtheta{2} \Pi_2}} \\
    & = \frac{\Pi_2 \Utheta{2}  \Pi_1 \Utheta{1}\rho_0 \Udagtheta{1} \Pi_1 \Udagtheta{2} \Pi_2}{p_2(\theta_1, \theta_2)}.
\end{align}
Note how the normalization constant of $\rho_1(\theta_1)$ cancels. Generalizing this to $M$ measurements, we get the general form
\begin{align}
    \rho_M(\theta_1,\ldots,\theta_M)& = \left(\overleftarrow{\prod^M_{m=1}} \Pi_m \Utheta{m}\right) \rho_0 \left(\overrightarrow{\prod^M_{m=1}}\Udagtheta{m} \Pi_m \right) p_{M}^{-1} (\theta_1, \ldots, \theta_M)\\
    & = \tilde{\rho}_M(\theta_1,\ldots,\theta_M)  p_M^{-1} (\theta_1, \ldots, \theta_M),
\end{align}
where 
\begin{align}
\tilde{\rho}_M (\theta_1, \ldots, \theta_M) &=  \left(\overleftarrow{\prod^M_{m=1}} \Pi_m \Utheta{m}\right) \rho_0 \left(\overrightarrow{\prod^M_{m=1}}\Udagtheta{m} \Pi_m \right),\\
    p_M (\theta_1, \ldots, \theta_M) &= \Tr{\tilde{\rho_M} (\theta_1, \ldots, \theta_M) },
\end{align}
are the unnormalized state and its normalization constant respectively.
We will omit the arguments $(\theta_1, \ldots, \theta_M)$ for now to reduce notational clutter.
We are interested in calculating gradients of a cost function that contains a sum of terms of the form $\Tr{\rho_M O}$ with respect to the parameter $\theta_l$, 
\begin{align}
    \dtheta{l} \Tr\{\rho_M O \} = \Tr{\left(\dtheta{l}  \tilde{\rho}_M \right) p_M^{-1} O} +  \Tr{\tilde{\rho}_M \left(\dtheta{l} p_M^{-1} \right)O}.
\end{align}
For the derivative of the unnormalized state, we get
\begin{align}
    \Tr{\left(\dtheta{l}  \tilde{\rho}_M\right) O}  & = \bra{\psi_0} \left(\overrightarrow{\prod^M_{m=1}}\Udagtheta{m} \Pi_m \right) O \left(\overleftarrow{\prod^M_{m=l+1}} \Pi_m \Utheta{m}\right) \Pi_l \dtheta{l}\Utheta{l} \left(\overleftarrow{\prod^{l-1}_{m=1}} \Pi_m \Utheta{m}\right) \ket{\psi_0}\\
    +& \bra{\psi_0}\left(\overrightarrow{\prod^{l-1}_{m=1}}\Udagtheta{m} \Pi_m \right) \dtheta{l} \Udagtheta{l} \Pi_l \left(\overrightarrow{\prod^M_{m=l+1}}\Udagtheta{m} \Pi_m \right) O \left(\overleftarrow{\prod^M_{m=1}} \Pi_m \Utheta{m}\right) \ket{\psi_0} \\
    & = \bra{\tilde{\psi}_0}\Udagtheta{l}\tilde{O} \dtheta{l}\Utheta{l} \ket{\tilde{\psi}_0} + \bra{\tilde{\psi}_0}\dtheta{l}\Udagtheta{l} \tilde{O} \Utheta{l}\ket{\tilde{\psi}_0},
\end{align} 
where
\begin{align}
    \ket{\tilde{\psi}_0} &=  \left(\overleftarrow{\prod^{l-1}_{m=1}} \Pi_m \Utheta{m}\right) \ket{\psi_0} \label{eq:psitilde},\\
    \tilde{O} &=\left(\overrightarrow{\prod^M_{m=l+1}}\Udagtheta{m} \Pi_m \right)O \left(\overleftarrow{\prod^M_{m=l+1}} \Pi_m \Utheta{m}\right).
\end{align}
If $U(\theta_l)$ is generated by a Pauli operator $A$, then $\dtheta{l}\Utheta{l} = -\frac{i}{2}A \Utheta{l}$ and so we can use the parameter-shift rule~\cite{Mitarai2018,Schuld2019}
\begin{align}
     -\frac{i}{2}\bra{\tilde{\psi}_0} \comm{A}{\Udagtheta{l} \tilde{O} \Utheta{l}}\ket{\tilde{\psi}_0}
    & = \frac{1}{2}\bigg(\bra{\tilde{\psi}_0}U^\dag(\theta_l+\frac{\pi}{2}) \tilde{O} U(\theta_l+\frac{\pi}{2}) -U^\dag(\theta_l-\frac{\pi}{2}) \tilde{O} U(\theta_l-\frac{\pi}{2}) \ket{\tilde{\psi}_0}\bigg)
\end{align}
If the state would be properly normalized, then this would provide a strategy for measuring this gradient. However, the final result
\begin{align}
    \Tr{\left(\dtheta{l}  \tilde{\rho}_M\right)p_M^{-1} O}  & =\frac{1}{2}\bigg(\bra{\tilde{\psi}_0}U^\dag(\theta_l+\frac{\pi}{2}) \tilde{O} U(\theta_l+\frac{\pi}{2}) - U^\dag(\theta_l-\frac{\pi}{2}) \tilde{O} U(\theta_l-\frac{\pi}{2}) \ket{\tilde{\psi}_0}\bigg) p_M^{-1},
\end{align}
calculates an observable with respect to an unnormalized state, because $p_M^{-1} $ normalizes the state without the parameter-shifted gates. Hence, we need to first normalize the state in order to be able to execute the gradient calculation on the device. We can achieve this by multiplying with the identity
\begin{align}
    \Tr{\left(\dtheta{l}  \tilde{\rho}_M\right) O} & =\frac{1}{2}\bra{\tilde{\psi}_0}\bigg(U^\dag(\theta_l+\frac{\pi}{2}) \tilde{O} U(\theta_l+\frac{\pi}{2})\times \frac{p_M^{+,l}(\theta_1, \ldots,\theta_l + \frac{\pi}{2}, \ldots, \theta_M)}{p_M^{+,l}(\theta_1, \ldots,\theta_l + \frac{\pi}{2}, \ldots, \theta_M)}\\
    &- U^\dag(\theta_l-\frac{\pi}{2}) \tilde{O} U(\theta_l-\frac{\pi}{2})\times \frac{p_M^{-,l}(\theta_1, \ldots,\theta_l - \frac{\pi}{2}, \ldots, \theta_M)}{p_M^{-,l}(\theta_1, \ldots,\theta_l - \frac{\pi}{2}, \ldots, \theta_M)}\bigg)  \ket{\tilde{\psi}_0}  p_M^{-1}  \\
    &  = \frac{1}{2} \left(\expval{O}^+ \frac{p_M^{+,l}}{p_M} - \expval{O}^- \frac{p_M^{-,l}}{p_M}  \right).\label{eq:unnormed_grad}
\end{align}
Here, $\expval{O}^\pm$ is the expectation value of the observable $O$ after the measurements $\{\Pi_1, \ldots, \Pi_M\}$ have been applied to the parameter-shifted circuit.\newline

\noindent For the gradient of the inverse of the normalization constant, we get
\begin{align}
    \Tr{\tilde{\rho}_M \left(\dtheta{l} p_M^{-1} \right)O} & = -\expval{O} p^{-1}_M \dtheta{l}p_M,
\end{align}
where we used the normalization constant to define $\expval{O}$, the expectation value of $O$ with respect to the measured circuit. The final step is to calculate $\dtheta{l} p_M$:
\begin{align}
    \dtheta{l} p_M &= \Tr{\dtheta{l} \tilde{\rho_M} (\theta_1, \ldots, \theta_M)} \\
    &= \bra{\psi_0} \left(\overrightarrow{\prod^{M-1}_{m=1}}\Udagtheta{m} \Pi_m \right) \Udagtheta{M} \Pi_M \Utheta{M} \left(\overleftarrow{\prod^{M-1}_{m=l+1}} \Pi_m \Utheta{m}\right) \Pi_l \dtheta{l}\Utheta{l} \left(\overleftarrow{\prod^{l-1}_{m=1}} \Pi_m \Utheta{m}\right) \ket{\psi_0}\\
    +& \bra{\psi_0}\left(\overrightarrow{\prod^{l-1}_{m=1}}\Udagtheta{m} \Pi_m \right) \dtheta{l} \Udagtheta{l} \Pi_l \left(\overrightarrow{\prod^{M-1}_{m=l+1}}\Udagtheta{m} \Pi_m \right) \Udagtheta{M} \Pi_M \Utheta{M} \left(\overleftarrow{\prod^{M-1}_{m=1}} \Pi_m \Utheta{m}\right) \ket{\psi_0} \\
    & = \bra{\tilde{\psi}_0}\Udagtheta{l} \tilde{\Pi}_M \dtheta{l}\Utheta{l} \ket{\tilde{\psi}_0} + \bra{\tilde{\psi}_0}\dtheta{l}\Udagtheta{l} \tilde{\Pi}_M \Utheta{l}\ket{\tilde{\psi}_0}
\end{align}
where 
\begin{align}
     \tilde{\Pi}_M &= \left(\overrightarrow{\prod^{M-1}_{m=l+1}}\Udagtheta{m} \Pi_m \right) \Udagtheta{M} \Pi_M \Utheta{M} \left(\overleftarrow{\prod^{M-1}_{m=l+1}} \Pi_m \Utheta{m}\right),
\end{align}
and $\ket{\tilde{\psi}_0}$ is the same as in~\cref{eq:psitilde}.
Again, we can apply the parameter-shift rule to obtain
\begin{align}
    \dtheta{l} p_M & = \frac{1}{2}\bigg(\bra{\tilde{\psi}_0}U^\dag(\theta_l+\frac{\pi}{2}) \tilde{\Pi}_M  U(\theta_l+\frac{\pi}{2}) - U^\dag(\theta_l-\frac{\pi}{2}) \tilde{\Pi}_M  U(\theta_l-\frac{\pi}{2}) \ket{\tilde{\psi}_0}\bigg).
\end{align}
But these expectation values are simply the normalization constants $p_M^{\pm,l}$, hence the final result becomes

\begin{align}
    \Tr{\tilde{\rho}_M \left(\dtheta{l} p_M^{-1} \right)O} = \expval{O} \frac{1}{2} \left(\frac{p_M^{+,l}}{p_M} -\frac{p_M^{-,l}}{p_M}  \right). 
\end{align}
All together, the final projective gradient is then
\begin{align}
    \dtheta{l} \Tr\{\rho_M O \} = \frac{1}{2} \left(\left(\expval{O}^+ + \expval{O}\right)\frac{p_M^{+,l}}{p_M} - \left(\expval{O}^- + \expval{O}\right)\frac{p_M^{-,l}}{p_M}\right).
\end{align}
We see that if all projectors are the identity projector, $\frac{p_M^{-,l}}{p_M}=1$, so we get back the old parameter shift rule
\begin{align}
     \dtheta{l} \Tr\{\rho_M O \} = \frac{1}{2}\left(\expval{O}^+ -\expval{O}^-\right).
\end{align}
Note that when we are measuring states, we are effectively creating a mixed quantum state, where each state in the mixture corresponds to a single projective state $\rho_M^{(i)}(\theta_1,\ldots,\theta_M)$, where $(i)=(i_1,\ldots, i_M)$, $i_j=0,1$ indicates a multi-index that determines the measurement outcomes in the computational basis (extensions to general measurements is trivial). The probability of the measurement $(i)$ occurring is given by the probability $p_M^{(i)} (\theta_1, \ldots, \theta_M)$. The total mixed state as a result of the measurements is then
\begin{align}
    \rho(\theta_1, \ldots, \theta_M) &= \sum_i p_M^{(i)} (\theta_1, \ldots, \theta_M)\rho_M^{(i)}(\theta_1,\ldots,\theta_M)\\
    &=  \sum_i \tilde{\rho}_M^{(i)}(\theta_1,\ldots,\theta_M)
\end{align}
From~\cref{eq:unnormed_grad} we see that the gradient of the mixed state is then simply
\begin{align}
    \Tr{(\partial_{\theta_l}\rho )O}& = \sum_i^{2^M} \frac{1}{2} \left(\expval{O}^{(i),+} \frac{p^{(i),+,l}_M}{p^{(i)}_M} - \expval{O}^{(i),-} \frac{p^{(i),-,l}_M}{p^{(i)}_M}  \right)
\end{align}
Hence the estimator for the gradient corresponds to the average expectation value over intermediate measurements done on parameter-shifted circuits weighted by $p^{(i),l}_M$. Therefore, the projective gradients can be estimated by obtaining statistics from the measurements done on the vanilla and parameter-shifted circuits. These statistics are obtained when estimating $\expval{O}$ and $\expval{O}^\pm$, respectively. 

\clearpage

\end{document}